\documentclass[conference]{IEEEtran}
\usepackage{cite}
\usepackage{graphicx}
\usepackage{textcomp}
\usepackage{url}

\usepackage{xspace}
\def\BibTeX{{\rm B\kern-.05em{\sc i\kern-.025em b}\kern-.08em
    T\kern-.1667em\lower.7ex\hbox{E}\kern-.125emX}}
    
\begin{document}

\title{Towards an interdisciplinary, socio-technical analysis of software ecosystem health
}

\author{\IEEEauthorblockN{Tom Mens}
\IEEEauthorblockA{\textit{Software Engineering Lab} \\
\textit{University of Mons}\\
Mons, Belgium \\
tom.mens@umons.ac.be}
\and
\IEEEauthorblockN{Bram Adams}
\IEEEauthorblockA{\textit{MCIS Lab} \\
\textit{Polytechnique Montreal}\\
Montr\'eal, Canada \\
bram.adams@polymtl.ca}
\and
\IEEEauthorblockN{Josianne Marsan}
\IEEEauthorblockA{\textit{Faculty of Business Administration} \\
\textit{Laval University}\\
Qu\'ebec, Canada \\
josianne.marsan@sio.ulaval.ca}
}

\maketitle

\newcommand{\secohealth}{\emph{SECOHealth}\xspace}

\begin{abstract}
This extended abstract presents the research goals and preliminary research results of the interdisciplinary research project \secohealth, an ongoing collaboration between research teams of Polytechnique Montreal (Canada), the University of Mons (Belgium) and Laval University (Canada). \secohealth aims to contribute to research and practice in software engineering by delivering a validated interdisciplinary scientific methodology and a catalog of guidelines and recommendation tools for improving software ecosystem health.
\end{abstract}

\begin{IEEEkeywords}
software, ecosystem, evolution, health, recommendation, prediction, survival, sustainability, resilience, socio-technical
\end{IEEEkeywords}

\section{Introduction}

The two-year interuniversity \secohealth project\footnote{\url{www.secohealth.org}} started on October 1, 2017. The three authors of this extended abstract are its Principal Investigators.
\secohealth aims to contribute to research and practice in software engineering by delivering an interdisciplinary scientific methodology and a catalog of guidelines and recommendation tools for improving software ecosystem (SECO) health. Those will enable key ecosystem actors to better monitor and control the SECO health and equip them with corrective and preventive measures to ensure their SECO's survival and sustainability. The interdisciplinary methodology used in our project will also guide other researchers in interdisciplinary projects involving open source communities or SECOs.

SECOs are large collections of interacting and interdependent software projects that share a common technological platform and that are maintained by large online communities of contributors. They pervade every aspect of human life including entertainment, health, economy, industry, politics, education and science. Commercial SECOs such as mobile app stores or the Internet-of-Things have taken over our daily lives by storm, to the extent that the functioning of our modern digitally-enabled society would be severely impacted if SECOs degrade in stability or even cease to exist.

Yet, despite the strategic importance of ensuring the overall well-being of SECOs, their health is still ill-apprehended, as SECOs are subject to constant evolution, due to an increasing pace of events (e.g., technological or environmental changes). What makes this especially challenging, is that SECOs do not have a centralised management for overseeing the ecosystem's health and survival. Instead, maintainers of SECO components need to understand and make decisions about the socio-technical impact of important events affecting SECO health and recommend corrective actions (e.g., improving SECO quality and its attractiveness to key actors). Unfortunately, there is only little support or best practices to enable SECO maintainers to perform these tasks.

\section{About software ecosystem health}

From a biological point of view, health can be defined as ``the extent to which an organism's vital systems are performing normally at any given time'' \cite{Wang2011ICIS}. This definition can be transposed to SECO health \cite{Jansen2014IST} by considering a SECO as a living organism, whose constituent software projects are the vital technical systems that need to perform ``normally'' in order to have a healthy ecosystem, and whose community is healthy if all community members are performing normally.

SECO health problems can be very diverse in nature, and can have many different causes. For example, in March 2016, the npm ecosystem experienced the problem of a package getting unpublished, causing several thousands of transitively dependent packages to break. The underlying cause was a typical case of rage quitting, where the owner of the package decided to remove all of its packages\footnote{\url{blog.npmjs.org/post/141577284765/kik-left-pad-and-npm}}.
Another documented example of rage quitting relates to ``toxic'' communication styles of open source communities, such as the one of the Linux Kernel community, causing a prominent developer to quit\footnote{\url{https://slashdot.org/story/15/10/05/2031247/linux-kernel-dev-sarah-sharp-quits-citing-brutal-communications-style}}; or the case where a central contributor to the bug handling community of Gentoo Linux unexpectedly left, causing a major disruption in the community's activity \cite{Zanetti2013CHaSE}.
From a more technical point of view, typical examples of health problems are packages containing bugs or security vulnerabilities, causing potential problems in packages depending on it. The impact of the problem grows as the number of transitive dependencies on a problematic package grows.

\section{Project Goals}

\secohealth aims at providing a scientific methodology and disciplined set of techniques to understand and control the health of software ecosystems. We adopt a socio-technical perspective since the technical and social layers of SECOs are strongly interwoven \cite{Mens2016ICSME}. Our project aims to: 
\begin{itemize}
\item define a conceptual model of SECO health;
\item explore analogies from other scientific disciplines such as ecology and toxicology;
\item determine indicators capable of measuring the different aspects of SECO health;
\item determine events that affect the health of a SECO and its constituent projects;
\item empirically validate these health indicators and events, both qualitatively and quantitatively;
\item build and evaluate models to predict the impact of a given event on SECO health;
\item build and evaluate a socio-technical dependency model to understand how health problems propagate throughout a SECO;
\item propose a catalog of guidelines and recommendations for supporting SECO health.
\end{itemize}

Joining our complementary strengths in theory-driven and data-driven investigation, we will follow a mixed-methods approach \cite{Johnson2004}, combining  bottom-up data mining and top-down interview/survey-based research, as well as combining state-of-the-art quantitative and qualitative analysis techniques emanating from different scientific disciplines.
 
Under the approval of Research Ethics Committees from the participating universities, we conducted face-to-face interviews 
at the European Open Source Summit of the Linux Foundation (Prague, October 2017) with 17 SECO practitioners.
The interviews followed the guidelines of Patton \cite{Patton2015}, with the goal of understanding what SECO health means for practitioners, what indicators they use themselves or could be used given the right data, and which events have impacted SECO health in the past.

We will operationalise the SECO health indicators into concrete metrics, and perform SECO data mining to measure and evaluate the identified health indicators. We will build and empirically validate prediction models of how SECOs will react to events, by relying on historical data from version control systems, code review and bug repositories, mailing lists and developer fora.

Based on the recent research on SECO and community health \cite{Fotrousi2014, Jansen2014IST, Monteith2014, ManikasK2015IWSECO}, we will consider three high-level characteristics of health: \emph{technical} (i.e., concerning technical software artefacts), \emph{social} (i.e., concerning contributor communities and the relations between their members) and \emph{phenomenological} (i.e., concerning external/internal events and their manifestation). Technical health characteristics include traditional software quality metrics, software dependency structure, software growth rate, size and frequency of software updates, bug fixes, security vulnerabilities, obsolete or deprecated components, and so on. Social characteristics include responsiveness of contributors (e.g., mean time to respond to a question, mean time to fix a bug), social network structure and its evolution (e.g. turnover rate), contributor activity and productivity, and the quality of interaction between all human stakeholders. Phenomenological health characteristics include the amount of company involvement (i.e., paid contributors), market share, presence of competing products, and so on.

With respect to the social health problem of developer turnover, we conducted an empirical study on the npm and RubyGems ecosystems. Using the statistical technique of survival analysis we identified which social or technical factors in a SECO coincide with a higher or lower probability of developer abandonment \cite{Constantinou2017ISSE}.

Concerning technical health, we carried out a quantitative empirical analysis of the evolution of package dependency networks for seven package distributions of varying size and age \cite{Decan2017EMSE}. We proposed metrics to capture the growth, changeability, reusability and fragility of these dependency networks. We observed that the dependency networks tend to grow over time, while a minority of packages are responsible for most of the package updates. The majority of packages depend on other packages, but only a small proportion of packages accounts for most of the reverse dependencies. We observed a high proportion of ``fragile'' packages due to a high and increasing number of transitive dependencies.

\section{Interdisciplinary Research}

\secohealth will view SECOs as ecological ecosystems comprised of a population of living organisms (interdependent software projects and their interacting communities of contributors), and  will produce health indicators and prediction models by drawing inspiration from well-known principles and theories from other disciplines, such as the notion of biodiversity in ecology \cite{Mens2014Chapter}, or the notion of toxicity in toxicology \cite{Carillo2016ICIS,Carillo2017AIS}. 

SECO health needs to be studied at different levels of granularity since the health of the SECO as a whole depends on the health of its social and technical components, and vice versa.
At a \emph{micro-level} of analysis (i.e., within and between individual projects of a SECO), we will explore the impact of \emph{toxicity}, arguing that certain behaviour and interactions in the SECO community can be toxic to not only the individual software projects, but even to the SECO as a whole, and hence can jeopardize its health and sustainability. Examples of possible toxic social behaviour may consist of deviant or aggressive behaviours, for example in the form of flame wars as a reaction to bad quality code contributions \cite{Carillo2016ICIS}. 
One promising way to assess such toxicity is by measuring \emph{social debt} \cite{Tamburri2015}, i.e., social interactions between SECO members that have been strained due to time pressure or lack of attention, and at some point might blow up and cause friction within the community of developers involved in a software project.

At a \emph{macro-level}, we will study how health problems of SECO components evolve and propagate to others. Among others, we will test the principle of \emph{biodiversity} by analysing to which extent the SECO's \emph{resilience} decreases when its diversity decreases. By resilience we refer to the ecosystem's capacity of resisting to disturbances, or recovering from a perturbation quickly. 
Diversity will be analysed according to a variety of factors (e.g., geographical, activity-related, time-related, gender-related, artefact-related). 
Some of these factors have been shown to have positive effects on software health. For example, gender diversity has been shown to have a positive effect on the productivity of GitHub teams \cite{Vasilescu2015CHI}.

\section{Related Work and Related Projects}

\secohealth can be considered as a successor project of the \emph{ECOS} project (Ecological Studies of Open Source Software Ecosystems) \cite{Mens2014ECOS} that was carried out from 2012 till 2017. As part of that project, we carried out multiple empirical studies of the evolution of open source software ecosystems (e.g., Gnome \cite{Vasilescu2014}, CRAN \cite{GermanAdamsHassan2013,ClaesMG2014,Decan2016SANER}, Debian \cite{Claes2015MSR}, npm and RubyGems \cite{Constantinou2016,Decan2017SANER,Constantinou2017}). 

The \secohealth members are actively involved in the CHAOSS (Community Health Analytics of Open Source Software) initiative of the Linux Foundation. Metrics Committee. Its goal is to define, implement and assess metrics for open source community health and sustainability. While CHAOSS' initial focus is on metrics at the level of individual software projects, \secohealth will focus on ecosystem-wide health metrics.
To operationalise our metrics, we aim to use Bitergia's  open source GrimoireLab tool chain for software development analytics\footnote{\url{http://grimoirelab.github.io}}, which is one of the tool suites actively supported by CHAOSS.

\section*{Acknowledgment}
This research is carried out in the context of FRQ-FNRS collaborative research project R.60.04.18.F ``SECOHealth'' and FNRS Research Credit J.0023.16 ``Analysis of Software Project Survival''. 
We thank our external project collaborators Kevin Carillo (Toulouse University, France), Damian Tamburri (Politecnico di Milano, Italy), Gregorio Robles (Universidad Rey Juan Carlos, Spain) and Bogdan Negoita (HEC Montr\'eal, Canada) for actively participating to this project.


\end{document}